\documentclass[final,3p,twocolumn,times]{elsarticle}

\usepackage{amssymb}
 \usepackage{amsthm}
\usepackage{cancel}

\usepackage{hyperref}
\usepackage{verbatim}

\usepackage{mathrsfs}
\usepackage{amsmath}
\usepackage{bm}
\usepackage{graphicx}
\usepackage{color}
\usepackage{multirow}

 \biboptions{square,sort&compress,comma}

\newcounter{bla}

\journal{Computer Physics Communications}

\begin{document}

\begin{frontmatter}



\title{A Simple and Fast Approach for Computing the Fusion Reactivities with Arbitrary Ion Velocity Distributions}




\author[label1,label2]{Huasheng XIE}
\ead{huashengxie@gmail.com, xiehuasheng@enn.cn}


\address[label1]{Hebei Key Laboratory of Compact Fusion, Langfang 065001, China}
\address[label2]{ENN Science and Technology Development Co., Ltd., Langfang 065001, China}

\begin{abstract}
Calculating fusion reactivity involves a complex six-dimensional integral of the fusion cross section and ion velocity distributions of two reactants. We demonstrate a simple Monte Carlo approach that efficiently computes this integral for arbitrary ion velocity distributions with a time complexity of $O(N)$, where $N$ is the number of samples. This approach generates random numbers that satisfy the reactant velocity distributions. In cases where these numbers are not readily available, we propose using Gaussian random numbers with weighted factors. For cases where only a small number of $N$ samples are available, a $O(N^2)$ method can be used. We benchmarked this approach against analytical results for drift bi-Maxwellian distributions and provided examples of drift ring beam and slowing down distributions. Our results show that the error can be less than 1\% with $N\sim10^4$ samples for our standard approach.

\end{abstract}

\begin{keyword}
Fusion Reactivity \sep Monte-Carlo \sep  Arbitrary Velocity Distributions  \\

\end{keyword}

\end{frontmatter}


\section{Introduction}

Fusion reactivity $\langle\sigma v\rangle$ is the integral of fusion cross section and the reactants' velocity distribution functions
\begin{equation}\label{eq:sgmv}
\langle\sigma v\rangle=\int\int d{\bm v}_1d{\bm v}_2\sigma(|{\bm v}_1-{\bm v}_2|)|{\bm v}_1-{\bm v}_2|f_1({\bm v}_1)f_2({\bm v}_2),
\end{equation}
where $f_1$ and $f_2$ are the normalized velocity distribution functions of two ions, i.e., $\int f_{j}({\bm v}_j)d{\bm v}_j=1$ with $j=1,2$, and $d{\bm v}_j=dv_{xj}dv_{yj}dv_{zj}$. Here, $\sigma=\sigma(E)$ or $\sigma=\sigma(v)$ is the fusion cross section, with $E$ being the energy in the center-of-mass frame
\begin{equation}
E=\frac{1}{2}m_rv^2,~~v=|{\bm v}|=|{\bm v}_1-{\bm v}_2|,~~m_r=\frac{m_1m_2}{m_1+m_2},
\end{equation}
where $m_1$ and $m_2$ are the mass of the two reactants, and $m_r$ is the reduced mass of the system.

Equation (\ref{eq:sgmv}) is not only important for calculating the fusion yield in laboratory \cite{Atzeni2004} or stellar \cite{Clayton1983} plasmas, but it is also useful for obtaining spectrum information of the distribution functions $f_{1,2}$ from a diagnostic perspective \cite{Appelbe2011}. However, calculating $\langle\sigma v\rangle$ for arbitrary $f_1$ and $f_2$ is difficult since it involves a six-dimensional (6D) velocity integral, which is usually computed numerically \cite{Lepage1978,Cordey1978}. Kolmes et al. \cite{Kolmes2021} used a mix of quadrature and Monte Carlo algorithm to study the fusion yield of plasma with velocity-space anisotropy at constant energy. Nath et al. \cite{Nath2013} reduced the 6D integral to a 3D integral for drift tri-Maxwellian distributions, which is numerically tractable. Several analytical 1D integral results are summarized in Ref. \cite{Xie2022}, with the drift bi-Maxwellian distribution being the most general one, which can be reduced to Maxwellian, bi-Maxwellian, and beam-Maxwellian cases.

Numerically integrating Eq.(\ref{eq:sgmv}) for arbitrary ion velocity distributions is generally considered to be complicated in the literature (see e.g. \cite{Nath2013}). Although a fast orthogonal polynomial expansion method was proposed in Ref. \cite{Cordey1978}, it is limited to velocity distributions that are independent of the azimuthal angle $\phi$ in spherical coordinates and therefore not generally applicable. Similarly, Ref. \cite{Appelbe2023} used a similar approach for energy spectra diagnostic of unscattered neutrons produced by deuterium-deuterium and deuterium-tritium fusion reactions. While Monte Carlo high-dimensional integral methods (see e.g. Ref. \cite{Lepage1978}) can be applied to arbitrary ion velocity distributions, their computation efficiency, i.e., computation speed and accuracy, highly depends on the sampling method used.

In this work, we propose a simple and effective Monte Carlo approach for computing the 6D integral in Eq.(\ref{eq:sgmv}). Unlike general Monte Carlo integral methods such as \cite{Lepage1978}, our approach is specifically designed for this problem, which enables us to achieve maximum computation efficiency. Moreover, we found that the approaches used in first-principle particle simulation codes \cite{Higginson2019,Wu2021} to calculate the fusion yield are valid for arbitrary velocity distributions and can be used to calculate Eq.(\ref{eq:sgmv}). However, our approach is more flexible when we are only interested in calculating the fusion reactivity integral in Eq.(\ref{eq:sgmv}). The proposed Monte Carlo approach has a time complexity of $O(N)$, where $N$ is the number of samples. In this paper, we also compare three types of this approach, which can be used for different situations.

Section \ref{sec:approach} describes the approach used in this work. In Section \ref{sec:apply}, we benchmark our results against analytical results for drift bi-Maxwellian distributions, and apply our approach to drift ring beam and slowing down distributions. Finally, in Section \ref{sec:summ}, we summarize our findings.

\begin{figure*}
\centering
\includegraphics[width=16cm]{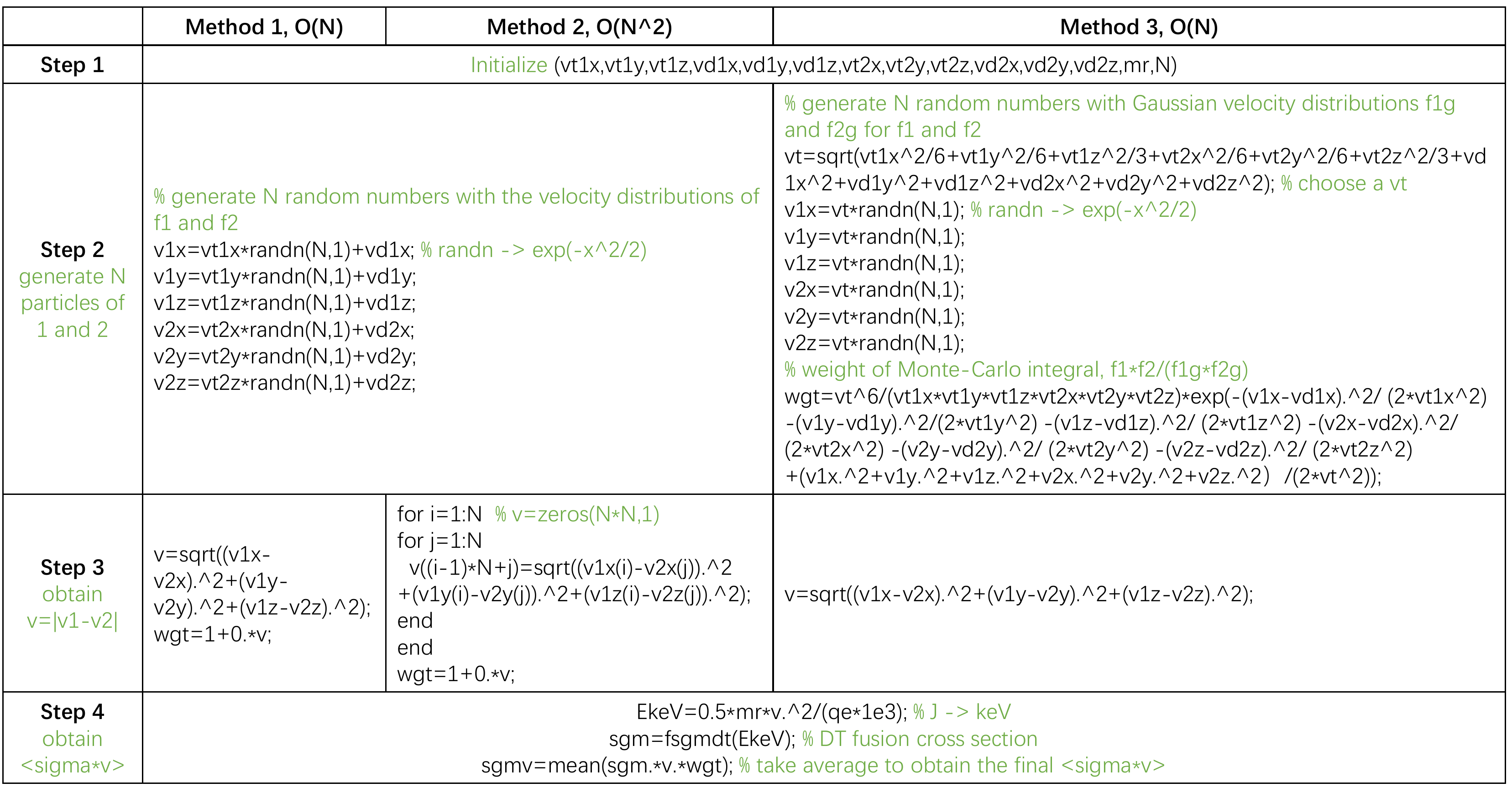}\\
\caption{Sample code demonstrating three Monte-Carlo methods for computing the 6D fusion reactivity integral for drift tri-Maxwellian velocity distributions.}\label{fig:cmp_method}
\end{figure*}

\section{Monte-Carlo Approach}\label{sec:approach}

The fusion reaction rate per unit volume and per unit time can be calculated as \cite{Atzeni2004,Clayton1983}
\begin{equation}\label{eq:R12}
R_{12}=\frac{n_1n_2}{1+\delta_{12}}\langle\sigma v\rangle,
\end{equation}
where $n_1$ and $n_2$ are the number densities of the two reactants, respectively, and $\delta_{12}$ is equal to 0 for different reactants and 1 for the same reactants.

Eq.(\ref{eq:R12}) implies a physical meaning, namely, that the fusion reactivity $\langle\sigma v\rangle$ represents the probability of a fusion reaction occurring. Thus, we select one particle from species 1 and one particle from species 2, and calculate $\sigma(|{\bm v}_1-{\bm v}_2|)|{\bm v}_1-{\bm v}_2|$ for these two particles. We repeat this process $N$ times, and as $N$ approaches infinity, the average value of each $\sigma(|{\bm v}_1-{\bm v}_2|)|{\bm v}_1-{\bm v}_2|$ will be the integral value of Eq.(\ref{eq:sgmv}). This yields a simple Monte-Carlo approach (Method 1, standard approach) to compute Eq.(\ref{eq:sgmv}):
\begin{itemize}
\item Step 1: Generate a random particle with velocity ${\bm v}_1=(v_{1x},v_{1y},v_{1z})$ that satisfies the velocity distribution $f_1({\bm v}_1)$, and a random particle with velocity ${\bm v}_2=(v_{2x},v_{2y},v_{2z})$ that satisfies the velocity distribution $f_2({\bm v}_2)$.
\item Step 2: Calculate $\sigma(|{\bm v}_1-{\bm v}_2|)|{\bm v}_1-{\bm v}_2|$ for these two particles.
\item Step 3: Repeat Steps 1 and 2 for $N$ times.
\item Step 4: Obtain the average value of each $\sigma(|{\bm v}_1-{\bm v}_2|)|{\bm v}_1-{\bm v}_2|$, which is the integral value of Eq.(\ref{eq:sgmv}).
\end{itemize}
This approach has a time complexity of $O(N)$.

In some situations, such as when using experimental diagnostic data, the number of samples $N$ may be small. In these cases, we can use the following approach (Method 2) to compute Eq.(\ref{eq:sgmv}):
\begin{itemize}
\item Step 1: Generate $N_1$ particles randomly with velocities ${\bm v}_1=(v_{1x},v_{1y},v_{1z})$ that satisfy the velocity distribution $f_1({\bm v}_1)$, and $N_2$ particles with velocities ${\bm v}_2=(v_{2x},v_{2y},v_{2z})$ that satisfy the velocity distribution $f_2({\bm v}_2)$.
\item Step 2: Calculate $\sigma(|{\bm v}_1-{\bm v}_2|)|{\bm v}_1-{\bm v}_2|$ for each pair of particles, resulting in a total of $N_1\times N_2$ pairs.
\item Step 3: Obtain the average value of each $\sigma(|{\bm v}_1-{\bm v}_2|)|{\bm v}_1-{\bm v}_2|$, which is the integral value of Eq.(\ref{eq:sgmv}).
\end{itemize}
Usually, $N=N_1\simeq N_2$. This approach has a time cost of $O(N_1N_2)\simeq O(N^2)$.

Both Method 1 and Method 2 require generating random numbers that satisfy the reactant velocity distributions. In cases where these numbers are not readily available, we can modify Method 1 to obtain Method 3, which uses weighted factors and the following equation
\begin{equation}\label{eq:sgmvw}
\langle\sigma v\rangle=\int\int d{\bm v}_1d{\bm v}_2\sigma(|{\bm v}_1-{\bm v}_2|)|{\bm v}_1-{\bm v}_2|w({\bm v}_1,{\bm v}_2)f_{1g}({\bm v}_1)f_{2g}({\bm v}_2).
\end{equation}
Here, the weight function is defined as
$$w({\bm v}_1,{\bm v}_2)=\frac{f_1({\bm v}_1)f_2({\bm v}_2)}{f_{1g}({\bm v}_{1})f_{2g}({\bm v}_2)}.$$
We can compute Eq. (\ref{eq:sgmvw}) using Method 3, which involves the following steps:
\begin{itemize}
\item Step 1: Generate a random particle with velocity ${\bm v}_1=(v_{1x},v_{1y},v_{1z})$ that satisfies the velocity distribution $f_{1g}({\bm v}_1)$, and another random particle with velocity ${\bm v}_2=(v_{2x},v_{2y},v_{2z})$ that satisfies the velocity distribution $f_{2g}({\bm v}_2)$.
\item Step 2: Calculate $\sigma(|{\bm v}_1-{\bm v}_2|)|{\bm v}_1-{\bm v}_2|w({\bm v}_1,{\bm v}_2)$ for these two particles.
\item Step 3: Repeat Steps 1 and 2 for $N$ times.
\item Step 4: Obtain the average value of each $\sigma(|{\bm v}_1-{\bm v}_2|)|{\bm v}_1-{\bm v}_2|w({\bm v}_1,{\bm v}_2)$, which is the integral value of Eq. (\ref{eq:sgmv}).
\end{itemize}
Method 3 is actually an important sampling Monte Carlo approach\cite{Lepage1978}. A good choice of $f_{1g}$ and $f_{2g}$ can reduce the requirement of $N$. In this work, we use Gaussian distributions for $f_{1g}$ and $f_{2g}$. The time cost of Method 3 is also $O(N)$.

Figure \ref{fig:cmp_method} provides sample code programs to demonstrate the above three Monte Carlo methods used to calculate the 6D fusion reactivity integral for drift tri-Maxwellian velocity distributions given by
\begin{eqnarray}\nonumber\label{eq:dtm}
&f_j({\bm v_j})=\Big(\frac{1}{2\pi}\Big)^{3/2}\Big(\frac{1}{v_{txj}v_{tyj}v_{tzj}}\Big)\exp\Big[-\frac{(v_{x j}-v_{dxj})^2}{2v_{tx j}^2}\\
&-\frac{(v_{y j}-v_{dyj})^2}{2v_{ty j}^2}-\frac{(v_{z j}-v_{dzj})^2}{2v_{tz j}^2}\Big].
\end{eqnarray}
Here, $v_{txj}$, $v_{tyj}$, and $v_{tzj}$ are the thermal velocities in each direction, and $v_{dxj}$, $v_{dyj}$, and $v_{dzj}$ are the drift velocities in each direction, with $j=1,2$. These three simple codes can quickly compute all the results in Nath et al \cite{Nath2013}, with Method 1 being the most effective (see Sec. \ref{sec:apply}).

\section{Benchmarks and Applications}\label{sec:apply}

To demonstrate the methods presented in Section \ref{sec:approach}, we compare the results with analytical solutions for drift bi-Maxwellian distributions\cite{Xie2022}. Additionally, we compare the three methods for drift ring beam\cite{Xie2019,Moseev2019} and slowing down\cite{Moseev2019} distributions and use the D-T fusion reaction cross-section data from Ref.\cite{Bosch1992}.

\begin{figure*}
\centering
\includegraphics[width=15cm]{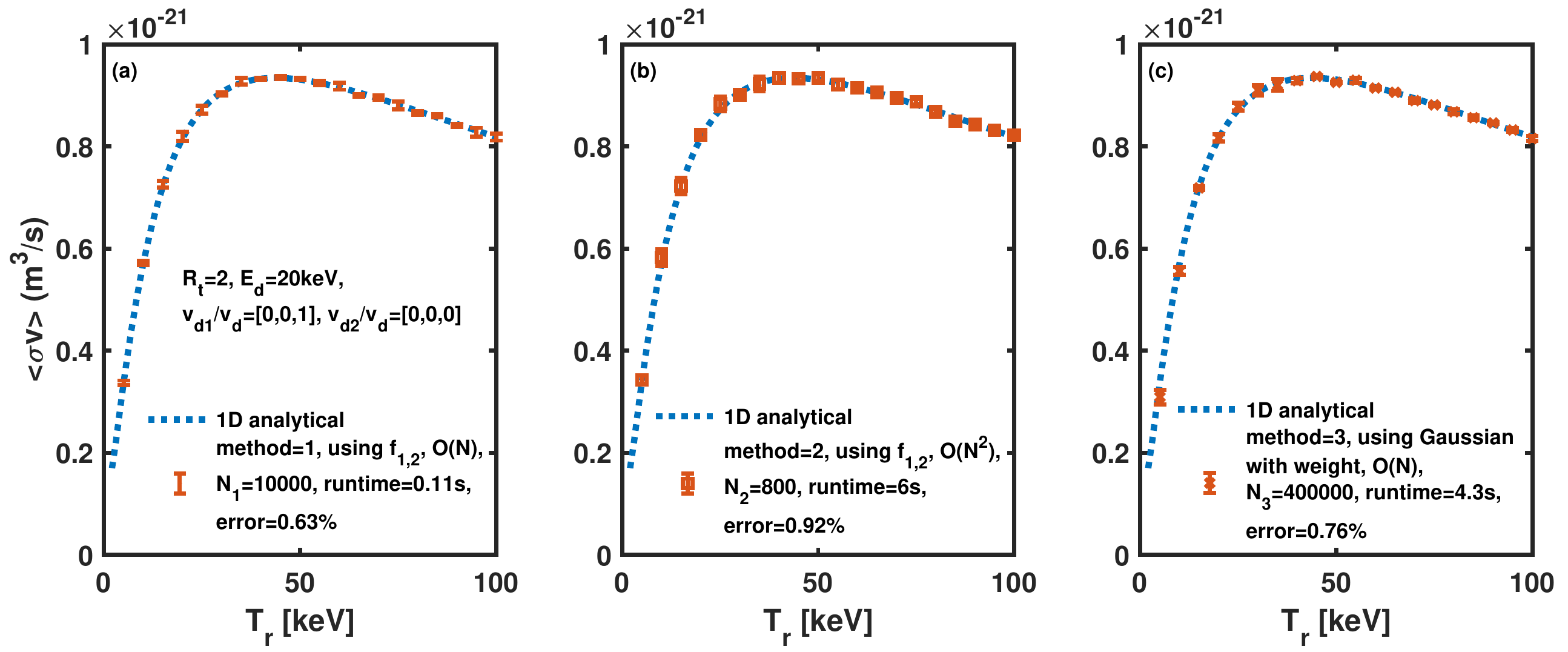}\\
\caption{Comparison between the results obtained using the 6D Monte-Carlo approach and the analytical 1D integral method\cite{Xie2022} for drift bi-Maxwellian distributions.}\label{fig:dbm1}
\end{figure*}

\begin{figure*}
\centering
\includegraphics[width=15cm]{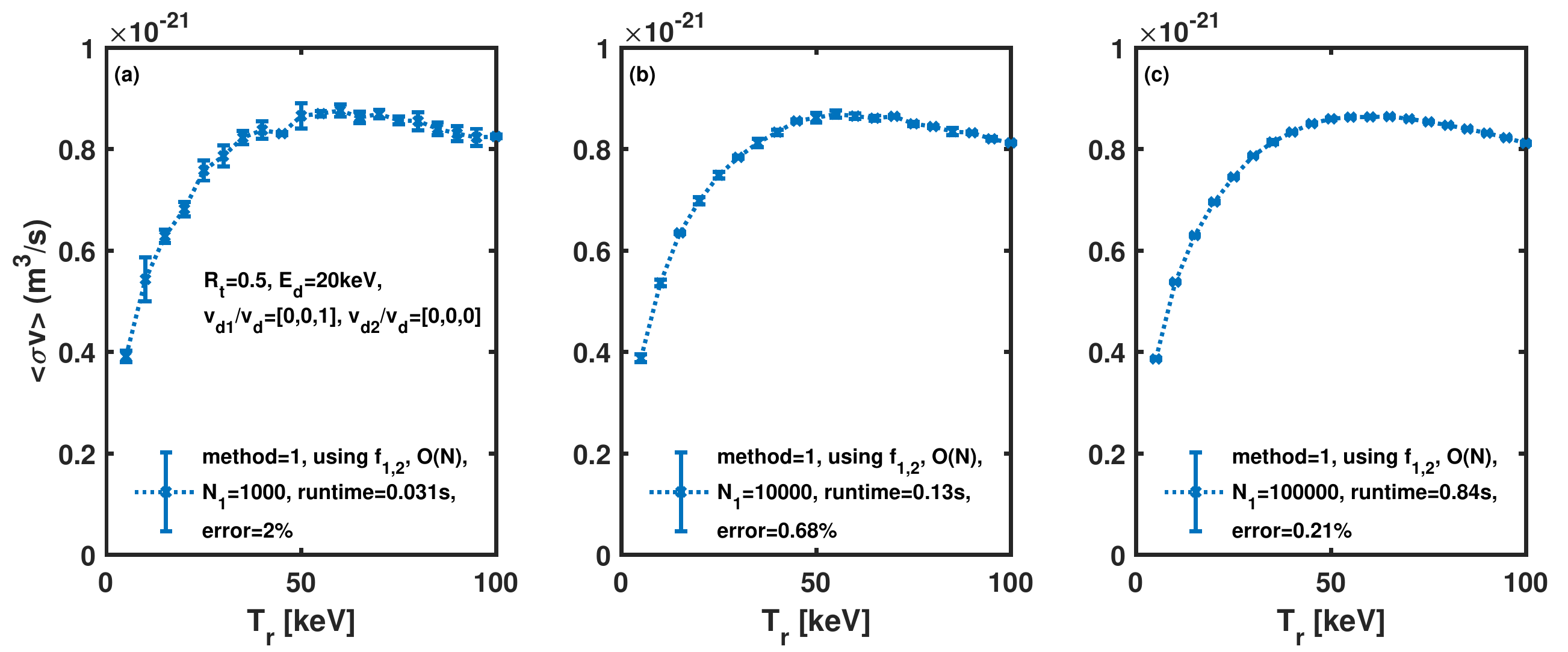}\\
\caption{Comparison of computation time and error using 6D Monte-Carlo approach Method 1 for drift bi-Maxwellian distributions with different values of $N$.}\label{fig:cmp_dtm_N}
\end{figure*}

\subsection{Drift bi-Maxwellian distribution}\label{sec:drb}
The distribution functions are given by
\begin{eqnarray}\nonumber\label{eq:dbm}
f_j({\bm v_j})&=&\frac{1}{T_{\parallel j}^{1/2}T_{\perp j}}\Big(\frac{m_j}{2\pi k_B}\Big)^{3/2}\cdot\\
&&\exp\Big[-\frac{m_jv_{\perp j}^2}{2k_BT_{\perp j}}-\frac{m_j(v_{\parallel j}-v_{dj})^2}{2k_BT_{\parallel j}}\Big],
\end{eqnarray}
where $j=1,2$, and $k_B$ is the Boltzmann constant. Here, $\int f_j({\bm v_j}) d{\bm v}_j=1$, $v_{\perp j}^2=v_{xj}^2+v_{yj}^2$, and $v_{\parallel j}=v_{zj}$. The drift tri-Maxwellian distribution Eq.(\ref{eq:dtm}) reduces to the drift bi-Maxwellian distribution Eq.(\ref{eq:dbm}) by taking $v_{txj}=v_{tyj}=\sqrt{k_BT_{\perp j}/m_j}$, $v_{tzj}=\sqrt{k_BT_{\parallel j}/m_j}$, and $v_{dxj}=v_{dyj}=0$ in Eq.(\ref{eq:dtm}). With this drift bi-Maxwellian distribution, the 6D integral Eq.(\ref{eq:sgmv}) reduces to a 1D integral\cite{Xie2022}, which is a function of only $T_r$, $R_t$, and $E_d$, where
\begin{eqnarray}\nonumber
T_r=\frac{(2T_{\perp r}+T_{\parallel r})}{3}, ~R_t=\frac{T_{\perp r}}{T_{\parallel r}},~E_d=k_BT_d=\frac{m_rv_{d}^2}{2},
\end{eqnarray}
where $v_{d}=v_{dz2}-v_{dz1}$. Additionally, we have
\begin{eqnarray}\nonumber
T_{\parallel r}=\frac{m_1T_{\parallel 2}+m_2T_{\parallel 1}}{m_1+m_2},~~T_{\perp r}=\frac{m_1T_{\perp 2}+m_2T_{\perp 1}}{m_1+m_2}.
\end{eqnarray}


Figure \ref{fig:dbm1} shows the benchmark results of the 6D Monte-Carlo approach to the analytical 1D integral\cite{Xie2022} for drift bi-Maxwellian distributions, which exhibit good agreement with $R_t=2$, $E_d=20$keV and $N_1=10^4$. To obtain the error of each method, the results are repeated three times for each case. The total computation time of each method is also the computer time taken. We observe that the total computer cost for computing the 6D Monte-Carlo results in Fig.\ref{fig:dbm1} using Method 1 for 20 points of $T_r$ with $N=10^4$ and repeat 3 times is 0.11 seconds, with an error less than 1\%. To achieve a similar level of accuracy, Methods 2 and 3 require around 50 times more computation time. Method 2 requires the smallest value of $N$ among these three methods.

Figure \ref{fig:cmp_dtm_N} compares the computation time and error with different values of $N$, using 6D Monte-Carlo approach Method 1 for drift bi-Maxwellian distributions. We find that $N=10^5$ is sufficient for these parameters ($R_t=0.5$, $E_d=20$keV). The computation time is not accurately proportional to $O(N)$ due to the fact that for high values of $N$, the vector program scheme can save some computation costs.

To make an accurate comparison of the performance of the three methods, it is necessary to use the same level of computational precision. However, since precision is influenced by many parameters and is difficult to control, this work can only provide a rough comparison. When the same level of computational precision is achieved, Method 1 requires a smaller number of samples, $N_1<N_2^2$, compared to Method 2. This is because Method 1 has a more accurate sampling of the distribution function than Method 2, while the calculation of the reactivity sum is similar. On the other hand, Method 3 requires a larger number of samples, $N_3>N_1$, compared to Method 1, which is understandable because Method 1 has a much simpler integral weight $\sigma(|{\bm v}_1-{\bm v}_2|)|{\bm v}_1-{\bm v}_2|$ than the weight $\sigma(|{\bm v}_1-{\bm v}_2|)|{\bm v}_1-{\bm v}_2|w({\bm v}_1,{\bm v}_2)$ in Method 3.

\begin{figure*}
\centering
\includegraphics[width=15cm]{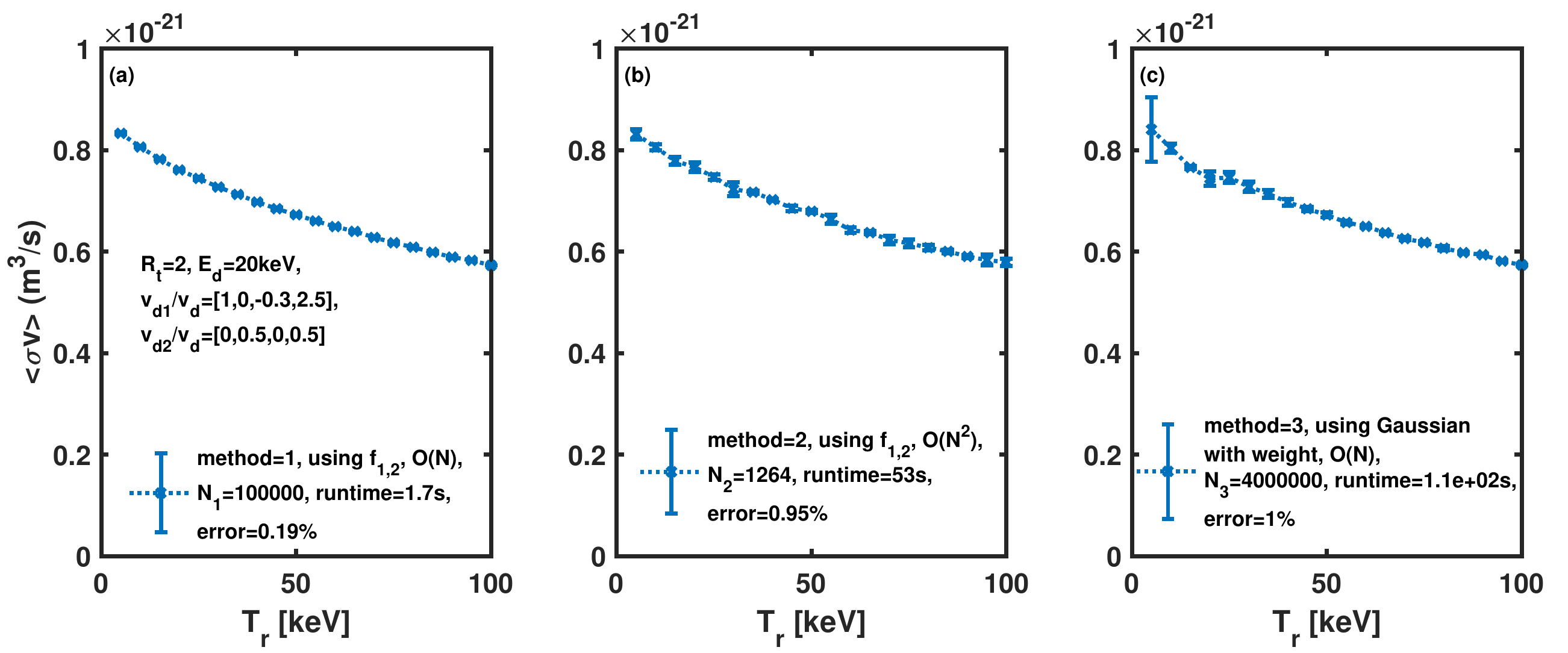}\\
\caption{Comparison of three Monte-Carlo methods for drift ring beam distributions, where $v_{dj}=[v_{djx},v_{djy},v_{djz},v_{djr}]$.}\label{fig:drm1}
\end{figure*}

\begin{figure*}
\centering
\includegraphics[width=15cm]{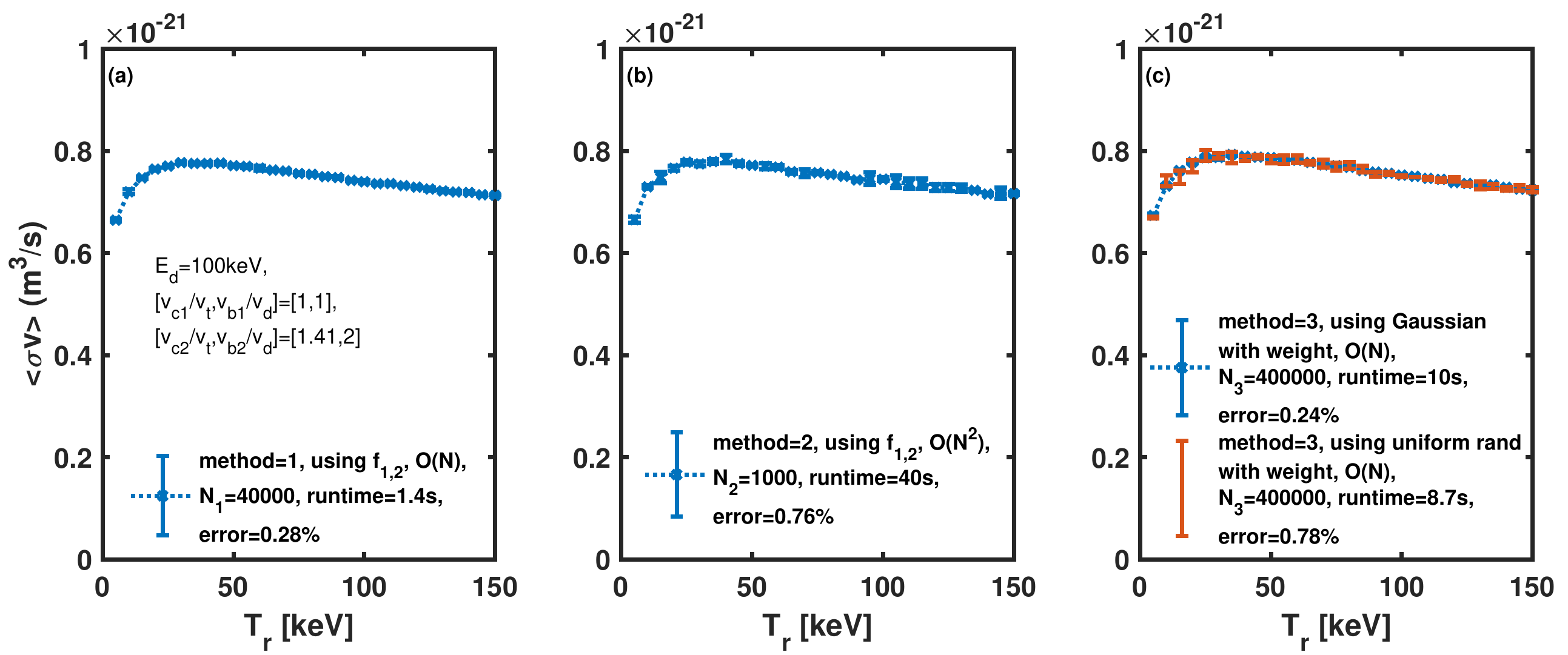}\\
\caption{Comparison of the three Monte-Carlo methods for slowing down distributions, where $v_t=\sqrt{2k_BT_r/m_r}$.}\label{fig:sd1}
\end{figure*}

\subsection{Drift ring beam distribution}\label{sec:drb}

The drift ring beam distribution, which includes both parallel and perpendicular drifts as well as temperature anisotropy, is given by\cite{Xie2019}
\begin{eqnarray}\nonumber\label{eq:drm}
f_j({\bm v_j})&=f_{zj}\cdot f_{\perp j}=\frac{1}{\sqrt{\pi} v_{tzj}}\exp\Big[-\frac{(v_{z j}-v_{dzj})^2}{v_{tz j}^2}\Big]\cdot \\
&\frac{1}{\pi A_j v_{t\perp j}^2}\exp\Big[-\frac{(\sqrt{(v_{x j}-v_{dxj})^2+(v_{y j}-v_{dyj})^2}-v_{drj})^2}{v_{t\perp j}^2}\Big],
\end{eqnarray}
where $A_j=\exp(-\frac{v_{drj}^2}{v_{t\perp j}^2})+\sqrt{\pi} (\frac{v_{drj}}{ v_{t\perp j}}) {\rm erfc}(-\frac{v_{drj}}{v_{t\perp j}})$, and $\int f_{j}({\bm v}_j)d{\bm v}_j=1$, for $j=1,2$. The error function ${\rm erfc}(-x)=1+{\rm erf}(x)$, and ${\rm erf}(x)=\frac{2}{\sqrt{\pi}}\int_0^xe^{-t^2}dt$. The 1D analytical form of Eq.(\ref{eq:sgmv}) for this distribution is not yet available. Note also that there exists a $\sqrt{2}$ difference between the definition of thermal velocity $v_{t}$ here and Eq.(\ref{eq:dtm}).

\ref{sec:appendix} provides instructions on how to generate random numbers with this distribution. Figure \ref{fig:drm1} compares the drift ring beam using the three methods in Section \ref{sec:approach}. Once again, we observe that Method 1 is the most efficient among them, and Methods 2 requires smallest $N$.

\subsection{Slowing down distribution}\label{sec:sd}

The isotropic slowing down distribution is given by \cite{Moseev2019}
\begin{eqnarray}\label{eq:sd}
f_j({\bm v_j})=\frac{3}{4\pi\ln[1+{v_{bj}^3}/{v_{cj}^3}]}\frac{H(v_{bj}-v)}{v^3+v_{cj}^3},
\end{eqnarray}
where $\int f_{j}({\bm v}_j)d{\bm v}_j=1$ for $j=1,2$, and $H(x)$ is the Heaviside function, defined as $H(x<0)=0$, $H(x>0)=1$, and $H(0)=1/2$. The 1D analytical form of Eq.(\ref{eq:sgmv}) for this distribution is not yet available.

Instructions for generating random numbers with this distribution are provided in \ref{sec:appendix}. Figure \ref{fig:sd1} compares the slowing down distribution using the three methods described in Sec.\ref{sec:approach}. Once again, we see that Method 1 is the most effective. For Method 3, we also compared two types of random numbers: Gaussian $f_{1g,2g}$, and uniform $f_{1g,2g}$ in $v_{xj,yj,zj}\in[-v_{bj},v_{bj}]$ for $j=1,2$. Both methods yielded similar results, indicating the robustness of this approach.

\section{Summary and Discussion}\label{sec:summ}
We have developed a simple Monte-Carlo approach to compute the 6D fusion reactivity integral Eq.(\ref{eq:sgmv}) for arbitrary ion velocity distributions. We compared three types of this approach for several typical distributions, such as drift bi-Maxwellian, drift ring beam, and slowing down distributions. Our results show that this approach is both robust and effective.

The second method is suitable for situations when $N$ is small, with a time cost of $O(N^2)$. The first method is found to be the most effective one among them, with a time cost of $O(N)$. However, it still requires a routine to generate the corresponding random numbers of the given distributions, as in the second method. The third method uses a weight function to remove the requirement of generating corresponding random numbers, with a time cost of $O(N)$. For these three methods, the typical requirement for $N_{1,2,3}$ is $N_1\simeq10^4-10^{5}$, $N_2\simeq 5\sqrt{N_1}\simeq10^{3}$, and $N_3\simeq 50N_1\simeq10^{6}-10^{7}$.

Overall, our Monte-Carlo approach provides a practical and efficient tool for computing the fusion reactivity integral. Although the basic ideas behind our Monte Carlo approach for computing the fusion reactivity integral may not be new, the approach presented in this work is still noteworthy for its simplicity and efficiency. Similar Monte Carlo pairwise treatments have been used in particle simulation codes, such as those in Refs. \cite{Higginson2019, Wu2021}, to calculate the fusion yield for arbitrary velocity distributions. Furthermore, the Fokker-Planck binary collision model for plasma particle simulation \cite{Takizuka1977} can also be retrospectively related to this approach. However, our approach, as demonstrated with three methods, is more flexible and applicable to a wider range of situations where only the fusion reactivity integral, Eq.(\ref{eq:sgmv}), needs to be calculated. Thus, we believe that our approach is valuable and worth summarizing to the community. In future work, further optimization of the algorithms and exploring new applications of this approach in related fields can be pursued.  The computation source codes used in this work are avaiable at https://github.com/hsxie/fusionreactivity.

{\it Acknowledgments}
Discussions with Dong WU, Mu-zhi TAN, Ke LI and Feng WANG are acknowledged.

\appendix

\section{Random numbers for drift ring beam and slowing down distributions}\label{sec:appendix}
To generate a velocity $v$ with distribution $f(v)$ from a uniform $u\in[0,1)$ random number using a monotonic function transformation $v=v(u)$, we use the relation
\begin{equation}
v(u+\Delta u)=v+\Delta v, \Delta u=f(v)\Delta v,
\end{equation}
which can be written as
\begin{equation}
f(v)dv=du.
\end{equation}
Solving for $u$ gives
\begin{equation}
u=u(v)=\int f(v')dv'.
\end{equation}
We can then calculate the transformation $v=v(u)$ from the inverse function of $u=u(v)$.

To model the distributions of drift ring beams, we use the product of two distributions: $f(\bm{v}) = f_z(v_z) \cdot f_{\perp}(v_x,v_y)$, where $f_z(v_z)$ can be generated using a standard Gaussian random number function. The distribution $f_{\perp}(v_x,v_y)$ is given by
\begin{equation}\nonumber\label{eq:}
f_{\perp }=\frac{1}{\pi A v_{t\perp }^2}\exp\Big[-\frac{(\sqrt{(v_{x}-v_{dx})^2+(v_{y}-v_{dy})^2}-v_{dr})^2}{v_{t\perp}^2}\Big],
\end{equation}
where $v_\perp = \sqrt{(v_x-v_{dx})^2 + (v_y-v_{dy})^2} \in [0,\infty)$ and $\phi$ is the angle between the $x$-axis and the velocity vector $\bm{v}_{\perp}$ in the $xy$-plane. The quantity $A$ is defined as $A = \exp(-v_{dr}^2/v_{t\perp}^2) + \sqrt{\pi} (v_{dr} / v_{t\perp}) {\rm erfc}(-v_{dr}/v_{t\perp})$, and $\int f_j(\mathbf{v}_j)d\mathbf{v}_j = 1$.
In the $(v_\perp,\phi)$ space, we have $f(v_\perp,\phi) = f(v_\perp) f(\phi)$, where
\begin{eqnarray}\nonumber\label{eq:}
&f(v_\perp)=\frac{2v_\perp}{A v_{t\perp }^2}\exp\Big[-\frac{(v_\perp-v_{dr})^2}{v_{t\perp}^2}\Big],~0\leq v_\perp<\infty\\
&f(\phi)=\frac{1}{2\pi}, ~0\leq\phi<2\pi.
\end{eqnarray}
The coefficients are normalized such that $\int_0^{\infty}f(v_\perp)dv_\perp=1$ and $\int_0^{2\pi}f(\phi)d\phi=1$. To generate $\phi$, we use a uniform random number $u\in[0,1)$ and set $\phi=2\pi u$. 

The relationship between $v_\perp$ and the uniform random number $u$ is given by the following equation
\begin{eqnarray}\nonumber\label{eq:}
&u=\int f(v_\perp)dv_\perp=\frac{1}{A}\Big\{\sqrt{\pi}\frac{v_{dr}}{v_{t\perp}}\Big[{\rm erf}\Big(\frac{v_{dr}}{v_{t\perp}}\Big)-{\rm erf}\Big(\frac{v_{dr}-v_\perp}{v_{t\perp}}\Big)\Big]\\
&+\exp\Big(-\frac{v_{dr}^2}{v_{t\perp}^2}\Big)-\exp\Big(-\frac{(v_\perp-v_{dr})^2}{v_{t\perp}^2}\Big)\Big\},
\end{eqnarray}
which satisfies the requirements $u(0) = 0$ and $u(\infty) = 1$. In the case of a usual Maxwellian/Gaussian distribution with $v_{dr} = 0$ and $A = 1$, we have
\begin{eqnarray}\nonumber\label{eq:}
u=-\exp(-{v_\perp^2}/{v_{t\perp}^2})+1,
\end{eqnarray}
so that
\begin{eqnarray}\nonumber\label{eq:}
v_\perp=v_{t\perp}\sqrt{-\ln(1-u)},
\end{eqnarray}
which is one of the standard ways to generate a Gaussian random distribution. When $v_{dr}\neq0$, we can obtain the inverse function $v_\perp(u) = u^{-1}(v_\perp)$ numerically using 1D interpolation, since $u(v_\perp)$ is known and monotonically increasing. Then, we can obtain the velocity components $(v_x,v_y)$ using the following equations:
\begin{equation}\nonumber\label{eq:}
v_x = v_\perp\cos\phi + v_{dx},~ v_y = v_\perp\sin\phi + v_{dy}.
\end{equation}
Note that $\phi$ and $v_\perp$ should use independent random numbers $u$.

Similarly, for the slowing-down distribution in $(v, \phi, \theta)$ space, we have
\begin{eqnarray}\nonumber\label{eq:}
&f({\bm v})=f(v)f(\theta)f(\phi),~~f(\theta)=\frac{1}{\pi},~~f(\phi)=\frac{1}{2\pi},\\
&f(v)=\frac{3v^2}{\ln[1+{v_{b}^3}/{v_{c}^3}]}\frac{H(v_{b}-v)}{v^3+v_{c}^3},
\end{eqnarray}
which means $0\leq\theta<\pi$ and $0\leq\phi<2\pi$ are uniformly distributed. We have
\begin{eqnarray}\nonumber\label{eq:}
u=\int f(v)dv=\ln[1+{v^3}/{v_{c}^3}]/\ln[1+{v_{b}^3}/{v_{c}^3}],
\end{eqnarray}
with $v\in[0,v_b),~u\in[0,1)$, i.e.,
\begin{eqnarray}\nonumber\label{eq:}
v=v_c\Big\{\exp\Big[u\ln(1+{v_{b}^3}/{v_{c}^3})\Big]-1\Big\}^{1/3}.
\end{eqnarray}
After generating random numbers of $(v, \theta, \phi)$, we can obtain $(v_x, v_y, v_z)$ via
\begin{eqnarray}\nonumber\label{eq:}
v_x=v\sin\theta\cos\phi,~~v_y=v\sin\theta\sin\phi,~~v_z=v\cos\theta.
\end{eqnarray}

For arbitrary distributions, generating random numbers is not always straightforward. However, there are numerical libraries available, such as UNURAN \cite{UNURAN}.

\end{document}